\documentstyle[aps,twocolumn,epsf]{revtex}

\newcommand{\be}{\begin{eqnarray}}
\newcommand{\ee}{\end{eqnarray}}

\newcommand{\la}{\langle}
\newcommand{\ra}{\rangle}

\newcommand{\half}{ {\textstyle\frac{1}{2}} }

\renewcommand{\vec}{\bbox}

\def\dag {\dagger}
\def\del {\partial}
\def\bdel{\bar{\partial}}
\def \bd {\bar{\partial}}

\def\d {\delta}
\def\bz {\bar{z}}
\def\half {{\textstyle {1 \over 2}}}

\def\ra {\rangle}
\def\la {\langle}
\def\Tr {{\rm Tr}}
\def\bA {{\bar A}}
\def\bD {{\bar D}}
\def\bV {{\bar V}}
\def\bz {{\bar z}}

\def \C {{\cal C}}
\def \S {{\cal S}}
\def \vx {\vec{x}}
\def \vy {\vec{y}}

\begin{document}

\title{Mass Gap and Confinement in (2+1)-Dimensional Yang-Mills Theory}

\author{Dimitra Karabali}

\vspace{1in}

\address{Department of Physics, Rockefeller University, New York, NY 10021 \\
and \\
Department of Physics and Astronomy, Lehman College of the CUNY, New York, NY
10468}


\maketitle

\begin{abstract}

Using a gauge-invariant matrix parametrization of the gauge fields, we present an 
analysis of how the mass gap arises in (2+1)-dimensional Yang-Mills 
theory.
We further derive an analytical continuum expression for the vacuum
wavefunction and based on this we calculate the string tension which is in
excellent agreement with Monte Carlo simulations of the corresponding lattice
gauge theory.
\end{abstract}


\narrowtext


\section{Introduction}
\label{sec1}


While the perturbative
aspects of non-Abelian gauge theories were understood many years ago, 
nonperturbative
phenomena have been difficult to analyze. Detailed analyses and
calculational
techniques are still lacking, even though most of the qualitative features
are
more or less clear. 

In particular, the issue of confinement and generation of mass gap in
Quantum
Chromodynamics (QCD) still defies a clear quantitative analysis. We have
tried
to understand these nonperturbative mechanisms in a somewhat simpler but
still
physically relevant framework, the (2+1)-dimensional Yang-Mills theory. To
start
with we have neglected the quark degrees of freedom, keeping only the
gluons.
The confinement mechanism applies to gluons as well. 
The lowest excited states of the purely gluonic theory are not the
perturbative
gluons but rather massive, color-singlet bound states of gluons, the
so-called
glueballs. The lowest
glueball mass defines essentially the mass gap of the theory.

The study of non-Abelian gauge theories in a lower dimensional context,
such as
two spatial dimensions is interesting for at least two reasons. Primarily
it can
be a useful guide to the more realistic case of three dimensions; given
the
complexity of gauge theories, any further progress, even in a simpler
lower
dimensional context, should be interesting. Secondly, gauge theories in
two
spatial dimensions can be an approximation to the high temperature phase
of QCD
with the mass gap serving as the magnetic mass. It is the latter
connection that makes our work relevant in a thermal field theory context, 
although there is no explicit mention of temperature. 
For a resume of our work and the relation to (3+1)-dimensioanl magnetic mass, 
see V.P. Nair's contribution
to these proceedings.


\section{Hamiltonian analysis, gauge-invariant parametrization}
\label{sec2}

In an attempt to understand the origin of the mass gap and the confinement
mechanism, we have analyzed $SU(N)$ Yang-Mills theory in two
spatial dimensions in a Hamiltonian framework in terms of gauge-invariant
variables \cite{kara1,kara2,kara3}. One of the advantages of the Hamiltonian
approach (fixed time) 
is the fact that the theory is essentially
two-dimensional and
various mathematical techniques from two-dimensional conformal field
theory
become applicable and very useful. As is convenient for a Hamiltonian
formulation, we use the $A_0 =0$ gauge.  The gauge potentials are $A_i=
-it^aA^a_i,~i=1,2$, where
$t^a$ are hermitian $(N\times N)$-matrices which form a basis of the 
Lie algebra of $SU(N)$ with $[t^a,t^b]=if^{abc}t^c,~\Tr (t^at^b)= 
\half \delta^{ab}$. The Hamiltonian can be written as
\begin{mathletters}
\label{Hamiltonian}
\be
& {\cal H}= T+V \, , 
\\
& T= {e^2\over 2}\int E^a_iE^a_i\, ,~~~~~~~~~~~V= \frac{1}{2e^2} \int B^aB^a 
\ee
\end{mathletters}
where $e$ is the coupling constant, $E^a_i = -i \frac{\delta}{\delta
A_i^a}$ is the electric field and 
$B^a = \half \epsilon_{jk}(\partial_jA_k -\partial_kA_j +[A_j,A_k])^a$ is 
the magnetic field. In (2+1) dimensions, $e^2$ has the dimension of
mass.

Our basic strategy has been to reformulate the theory in terms of
gauge-invariant variables and analyze the corresponding Schr\"odinger
equation directly on the physical configuration space ${\cal {C}}
= {\cal{A}} / {\cal{G}}$ (space of gauge potentials modulo gauge transformations).
This is achieved by making use of a special parametrization of the gauge
fields in terms of complex matrices in the following way. We  combine the 
spatial coordinates
 $x_1 ,x_2$ into the complex combinations
$z=x_1 -ix_2,~{\bar z} =x_1+ix_2$; correspondingly we have
$A\equiv A_{z} = {1 \over 2} (A_1 +i A_2), ~~  {\bar A}\equiv A_{\bar{z}}
= {1
\over 2} (A_1 -i A_2) = - (A_z)^{\dagger}$. The parametrization we use is
given by
\be 
\label{param}
A_z = -\partial_{z} M M^{-1} \, ,~~~~~~~~~~~~~~~~ A_{\bar{z}} = M^{\dagger
-1}
\partial_ {\bar{z}} M^{\dagger} 
\ee
Here $M,~M^\dagger$ are complex $SL(N,{\bf {C}})$-matrices.  

From (\ref{param}) it is clear that the gauge transformation, $A_i
\rightarrow g A_i g^{-1} - \partial _i g g^{-1}$, is expressed in 
terms of $M,~M^{\dag}$ by
\be
M \rightarrow g M \, , ~~~~~~~~~M^{\dagger} \rightarrow M^{\dagger} g^{-1}
\ee
for $g(\vx) \in SU(N)$. In particular, if we split $M$ into a unitary part
$U$  and a
hermitian part $\rho$ as $M=U\rho$, then $U$ is the `gauge part' and it can
be removed by a gauge transformation. The combination $H=M^\dagger 
M =\rho^2$ is a 
gauge-invariant
field and it provides a parametrization for the configuration space $\C$. 
The hermitian field
$H$ belongs to
$SL(N,{\bf {C}})/SU(N)$.

The parametrization (\ref{param}) introduces a new symmetry in the
theory, what we call ``holomorphic" invariance. This arises as follows. The
original gauge fields $A$, $\bA$ remain invariant under the transformation
$M \rightarrow M \bV (\bz),~ M^{\dag} \rightarrow V(z) M^{\dag} , ~ H \rightarrow
V(z) H \bV (\bz)$, where $V,~\bV$ depend only on the holomorphic coordinate 
$z$ and the antiholomorphic coordinate $\bz$ respectively. Since the
original theory is unaffected by this transformation, we should require
that all physical states remain invariant under ``holomorphic"
tranformations.

Having derived a particular parametrization for ${\cal{C}}$ one is able to
ask more detailed questions regarding its structure. An important question
is: what is the integration measure (volume element) on ${\cal{C}}$ ? 
One
of the advantages of the particular parametrization that we
introduced is that we can give an explicit answer to the above
question.

The volume element in the original space of gauge potentials ${\cal{A}}$
is given by
\be
\label{measA}
d\mu ({\cal{A}})=\prod_{x,a} dA^a(\vx) dA^a (\vx) =\vert \det (D \bD ) \vert~ d\mu
(M,M^\dagger )
\ee
where $D,~ \bD$ are covariant derivatives in the adjoint representation;
$D = \del - \del M M^{-1},~\bD = \bdel + M^{\dag -1} \bdel
M^{\dag}$. $d\mu(M,M^\dagger )$ is the standard volume element in the space
of $SL(N,\bf{C})$-valued fields. Factorising $M$ into a unitary part $U$
and a hermitian part $\rho$, $M=U \rho$, $d\mu(M,M^\dagger )$ can be
written as
\be
\label{measM}
d\mu(M,M^\dagger ) = d\mu (H) d\mu (U)
\ee
where $d\mu(U)$ is the Haar measure for $SU(N)$ and $d\mu(H)$ the
corresponding measure for hermitian fields. The volume element for
${\cal{C}}$ is now obtained as
\be
\label{measC}
d\mu (\C )~= {d\mu
({\cal{A}}) \over d\mu (U)} = \vert \det (D \bD)\vert d\mu(H)
\ee
The problem is thus reduced to calculating the determinant of the
two-dimensional operator $D\bD$. Using a gauge-invariant regulator we find
\cite{WZW,correlators}
\be
\label{meas}
d\mu ({\cal{C}}) \sim e^{2c_A \S(H)}~d\mu(H)
\ee
where $c_A$ is the adjoint quadratic Casimir ($c_A=N$ for $SU(N)$) and
${\cal{S}}(H)$ is the Wess-Zumino-Witten (WZW) action for the hermitian
matrix field $H$ given by
\be
&&\S (H)  =  {1 \over {2 \pi}} \Tr \int d^2x~\partial H \bar{\partial} H^{-1}
\nonumber
\\ 
& + & {i
\over {12 \pi}} \Tr \int d^3x \epsilon ^{\mu \nu \alpha} H^{-1} \partial
_{\mu} H
H^{-1}
\partial _{\nu}H H^{-1} \partial _{\alpha}H 
\ee
As is typical for the WZW action, the second integral is over a three-dimensional
space $M^3$ whose boundary is the physical two-dimensional space corresponding to
the coordinates $z,{\bar z}$. The integrand thus requires an extension
of the matrix field $H$ into the interior of $M^3$, but physical results do not depend
on how this extension is done \cite{witten}. Actually for the special case of hermitian
matrices, the second term can also be written as an integral over spatial
coordinates only \cite{correlators}.

The inner product for gauge-invariant physical states is given by
\be
\label{inner}
\la 1 \vert 2\ra = \int  d\mu (H)  e^{2c_A \S (H)}~\Psi_1^* (H) \Psi_2 (H)
\ee
Here we begin to see how conformal field theory could be useful; equation
(\ref{inner}) shows that matrix elements of the (2+1)-dimensional gauge theory
can
be thought of as correlators of the hermitian WZW model, which is a
conformal field theory. This point of view allows us to derive further
constraints on the dependence of the wavefunction on the gauge-invariant
variables $H$. The correlators of the hermitian WZW model have been studied 
before \cite{correlators} and following conformal field theoretic arguments one
can show that
all wavefunctions of finite inner product are functions of the current
\be
\label{current}
J_a = {c_A \over \pi} (\del H H^{-1})_a
\ee
The result is not surprising. The Wilson loop operator itself can be written
in terms of the current as
\be
W(C)~=~ \Tr ~P ~e^{-\oint_C (Adz+\bA d{\bar z})}~=
 \Tr ~P~e^{(\pi /c_A)\oint_C J }
\ee
Since in principle all gauge-invariant functions of $A, \bA$ can be
constructed from $W(C)$, the current $J$ should suffice to generate all the
gauge-invariant states of the theory. It is thus important to express the
original Hamiltonian in terms of the current $J$, which can be thought of 
as a collective field.

The potential energy term can be written as 
\be\label{potential}
V = {1 \over {2 e^2}} \int B^2 (\vx) = { \pi \over {m c_A}} \int \bdel J_a
(\vx)
\bdel J_a (\vx)
\ee
where $m= \frac{e^2 c_A}{2 \pi}$.

The kinetic energy term becomes
\be
\label{kinetic}
T= & m & \left[ \int_x J^a(\vx) {\d \over \d J^a(\vx)} + \int
\Omega_{ab}
(\vx,\vy) 
{\d \over \d J^a(\vx) }{\d \over \d J^b(\vy) }\right] \nonumber
\\
& \Omega & _{ab}(\vx,\vy)= {c_A\over \pi^2} {\d_{ab} \over (x-y)^2} ~-~ 
i {f_{abc} J^c (\vy)\over {\pi (x-y)}} \, 
\ee
The first term in $T$ essentially counts the number of $J$'s and the second
term replaces pairs of $J$'s by the operator product expansion for currents
in the WZW-model. It is the first term which is responsible for the
appearance of a mass gap. As it is obvious from (\ref{kinetic}), the lowest
excited eigenstate of $T$ is the current $J$ itself:
\be
\label{gap}
T J_a(\vx) = m J_a (\vx)
\ee
The current $J$ can be thought of as the nonperturbative gauge-invariant
gluon with dynamical mass $m$. Although $J_a$ is an eigenstate of $T$, it is
not an acceptable physical state. As we mentioned earlier, physical states
should be invariant under ``holomorphic" transformations. The state with the
minimum number of $J$'s which satisfies ``holomorphic" invariance is a
$2J$-state of the form $\bdel J_a \bdel J_a$. This is also a color-singlet
state as expected. Such an expression after a careful regularization
(``holomorphic" invariant point splitting) is a good candidate for the
lowest glueball state $O^{++}$
\be
\Phi_2 =\int f(\vx, \vy) \Tr : \bdel J (\vx) \bigl(H(x,\bar{y}) H^{-1}
(y,\bar{y})
\bigr) \bdel J (\vy):
\ee
Such a state has a mass gap of at least $2m$. The regulator term introduces
interactions among currents which make the estimation of the binding energy
rather difficult. 

\section{Vacuum Wavefunction and String Tension} 
\label{sec3}

So far we have considered the spectrum of the kinetic energy term $T$.
Notice that the potential energy term $V$, (\ref{potential}), gives
contributions of the order
$k^2 / e^2 \sim k^2 / m$, where $k$ is a typical momentum. 
For momentum modes $k \ll e^2 \sim m $, the kinetic energy term is dominant
and its spectrum provides an approximation for the spectrum of the whole
theory. In this limit, which can be thought of as the strong coupling limit,
the potential energy term $V$ can
be treated in a perturbative fashion. Let us see how this
works in the calculation of the vacuum wavefunction.

As far as the kinetic
energy operator is concerned, $\Phi_0 =1$ may be taken as the vacuum
wavefunction. This might seem rather trivial, however it is important to notice
that the corresponding norm $\la \Phi_0 | \Phi_0 \ra$, which is also the volume of
the configuration space ${\cal{C}}$, coincides with the partition function
of the hermitian
WZW-model, which upon appropriate regularization is finite \cite{correlators}. In
other words $\Phi_0$ is normalizable with the 
inner product (\ref{inner}). Essentially the exponential factor in (\ref{inner})
provides the necessary damping to render the norm finite. (It is infinite in
the Abelian case $c_A =0$, where the exponential factor is absent.)

For low momentum modes, $k \ll m $ , the inclusion of the potential energy
term leads 
to a modified vacuum wavefunction which can be written as
\be
\Psi_0 = e^P \Phi _0
\ee
where $P$ is a functional of the $J$'s which 
can be expanded in powers of $1/m$.
The various terms in this expansion can be determined from the Schr\"odinger
equation
\be
\label{Schrod}
{\cal H} \Psi_0 = (T +V) \Psi_0 =0
\ee

The first few terms are given as \cite{kara3}
\be
\label{expansion}
 P =   &-& {\pi \over { m^2 c_A}} \Tr \int  : \bdel J \bdel J : \nonumber
\\
& - & \left({\pi \over { m^2 c_A}}\right)^2 \Tr \int   \bigl[: \bdel J ( {\cal
D} \bd ) \bdel J 
    +  {1 \over 3} \bdel J  [J, \bdel ^2 J] : \bigr] \nonumber
    \\  
& - & 2 \left({\pi \over {m^2 c_A}}\right)^3 \Tr \int \bigl[ : \bdel J  (
{\cal D} \bd )^2 \bd J 
+{2 \over 9} [ {\cal D} \bd J,~ \bd J] \bd ^2 J  \nonumber
\\
& + & {8 \over 9} [{\cal D} \bd
^2
J,~ J] \bd ^2 J - {1 \over 6} [J, ~ \bd J] [\bd J,~ \bd ^2 J] \nonumber
\\
& - & {2 \over 9}
[J,
\bd J][J, \bd ^3 J]: \bigr]
~ +  {\cal O} ( {1 \over m^8})  
\ee
where ${\cal D}h= {c_A \over \pi} \del h -[J,h]$. The normal ordering of
various
terms in (\ref{expansion}) is necessary for $P$ to satisfy (\ref{Schrod}). 
The second derivative term
in (\ref{kinetic}) can give singularities when acting on composite operators.
The
normal
ordering subtracts out precisely these singularities.

Our approach provides the first analytical, continuum calculation of the
vacuum wavefunction. There are several interesting points regarding this. 
The leading order
term for the vacuum wavefunction is
\be
\label{leading}
\Psi_0 & \approx & \exp \left[ -{\pi \over { m^2 c_A}} \Tr \int :\bdel J \bdel J :
\right] \nonumber \\
& = & \exp \left[ -{1\over 2me^2} \Tr \int B^2\right] 
\ee
The calculation of expectation values involves averaging with the
factor $\Psi_0^* \Psi_0 \approx e^{-S}$, where $S$, as seen from the above
equation, is the action of a Euclidean two-dimensional Yang-Mills
theory of coupling constant $g^2 = m e^2 = e^4 c_A /2 \pi $. 
Thus, retaining only the leading term in $\Psi_0$, 
the expectation value of the Wilson loop 
in the fundamental representation is given by \cite{Wloop}
\be
\label{Wilson}
\la W_F (C) \ra = \exp \left[ - {e^4 c_A c_F \over 4\pi } {\cal
A}_C \right]
\ee
where ${\cal A}_C$ is the area of the loop $C$ and $c_F$ is the quadratic
Casimir of the
fundamental representation. The expectation value of the Wilson loop
exhibits 
an area law behavior, as expected for a confining theory. 
From (\ref{Wilson}) we can identify the expression for
the string tension $\sigma $ as
\be
\label{tension}
\sigma = {e^4 c_A c_F \over 4\pi } = e^4 \left( {N^2 -1\over 8\pi }\right)
\ee

Recent Monte Carlo calculations of the string tension give the values
$\sqrt{\sigma}/e^2 =$ 0.335, 0.553, 0.758, 0.966  for the gauge 
groups $SU(2),~SU(3),
~SU(4)$ and $SU(5)$
respectively \cite{teper}. The corresponding values calculated from
(\ref{tension}) are
0.345, 0.564, 0.772, 0.977. 
We see that there is excellent agreement (upto $\sim 3\%$) 
between (\ref{tension})
and the Monte Carlo results. It is further interesting to notice that our
analytic expression for the string tension (\ref{tension}) has the appropriate
$N$-dependence as expected from large-$N$ calculations. 

Another interesting point is the fact that we have obtained an expression
(\ref{expansion})
local in terms of the current $J$ but nonlocal in terms of the original
magnetic field $B$. 
All previous attempts to derive an analytical expression of the vacuum
wavefunction are based on the assumption that, in the strong coupling regime, such
a wavefunction can be written as a local expansion in $B$ \cite{conj,lattice}. 
Our results obviously do not agree with this approach. It
is thus interesting to investigate whether one could incorporate these nonlocal
terms and verify our results in lattice calculations.

The approximation of the vacuum wavefunction by the first few terms in
(\ref{expansion}) makes sense
only in the low momentum regime $k \ll m$. 
Although very cumbersome, in principle we could determine arbitrarily many terms in
the series expansion (\ref{expansion}), so one would expect that by resumming up
various terms one might be able to get information on the vacuum wavefunction away
from the low momentum region.
In particular, what about the short distance behavior, $k \gg e^2$ ? Indeed this 
can be retrieved by converting the $1/m$
expansion into
a series expansion in $J$'s. The terms in (\ref{expansion}) can be naturally 
rearranged into terms with
two
$J$'s, terms with three $J$'s, etc. The series of terms (infinite many) with
only two $J$'s can be summed up to give
\be
\label{2J}
P = -{1 \over {2 e^2}} \int_{x,y} B_a(\vx) \left[{ 1 \over  {\bigl( m + 
\sqrt{m^2 - \nabla ^2 } \bigr)} }\right] _{\vx,\vy} B_a(\vy)
\ee
In the weak coupling or high momentum
regime $k \gg e^2$, the leading order term in (\ref{2J}) is
\be
P = - {1 \over {2 e^2}} \int_{x,y}~ B_a(\vx) \left[ 
(-\nabla^2)^{-\half}\right]_{\vx,\vy}  B_a(\vy)
\ee
This is the vacuum wavefunction for an Abelian theory as expected.
Of course, in the low momentum regime $k \ll e^2$ the 
leading order term in 
(\ref{2J}) will
reproduce (\ref{leading}). 

Now what about the 3$J$-terms and higher order terms? Part of the
$3J$-contribution is just what is needed to covariantize the Laplacian operator
$\nabla^2$ in (\ref{2J}), but that is not all. We can explicitly derive the full 
$3J$-contribution \cite{kara3}. What is important is the fact that comparing the
$2J$ and $3J$-terms we find \cite{kara3} that the $3J$-term is subdominant
compared to the leading $2J$-term in $P$ for both the low and high momentum
regimes. Similar arguments hold, based on dimensional analysis, for the higher
$J$-terms. So the expansion of $P$ as a series expansion in $J$'s is consistent
with both low and high momentum regimes.

So far we have discussed the structure of the vacuum wavefunction. One
could, in principle, 
extend the above analysis for
the excitation spectrum of the Schr\"odinger equation. For example,
in the absence of the 
potential term $V$,
the current $J$ is an eigenstate of the kinetic energy operator $T$,
(\ref{gap}), with
eigenvalue $m$. One could now ask what the corresponding modified 
eigenstate and eigenvalue should be, once the
potential term is included. We find
\be
(T + V) J_a ~ e ^P =  \sqrt{m^2 - \nabla ^2} ~J_a ~ e ^P
\ee
where in the expression for $P$ we have kept only the leading $2J$-term
contribution and
have neglected the higher $J$-terms.
As expected in a relativistic theory, the mass $m$ gets corrected to its 
relativistic
expression $\sqrt{m^2 + \vec{k} ^2}$, which also displays the correct high and low
momentum limits.  In a similar fashion we can obtain
expressions interpolating between short and long distance regimes for other
quantities of interest.
\vskip .1in
\acknowledgements

The talk presented here is based on work done in collaboration with C. Kim and 
V.P. Nair.
It was supported in part by the DOE grant DE-FG02-91ER40651-Task B.
Special thanks to Prof. U. Heinz and the other organisers of the ``5th
International
Workshop of Thermal Field Theories and Their Applications" for an interesting and
very well organized meeting.

\vskip .1in


\end{document}